\documentclass[a4paper,11pt]{article}
\usepackage[DIV12]{typearea}

\usepackage[latin1]{inputenc}
\usepackage[english]{babel}
\usepackage{amsfonts}
\usepackage{amsmath}    %NOTE: \beq, \eeq abbrev. won't work
\usepackage{amssymb}
\usepackage{cite}
\usepackage{mathrsfs}
\usepackage{pifont}
\usepackage{units}
\usepackage[small,loose]{subfigure}  % subfigures with (a), (b) etc., ... and own caption
\usepackage{cancel}
\usepackage[pdftex]{color}
\usepackage{slashed}
\usepackage{graphicx}

\usepackage[pdftex,colorlinks,pdfpagelabels]{hyperref}
\usepackage[figure,table]{hypcap} % Correct a problem with hyperref
\hypersetup{
   bookmarksnumbered,
   pdfstartview={FitV},
   pdfpagemode={UseOutlines},
   pdfauthor={Jasper Hasenkamp and Martin Winkler},
   pdftitle={},
   pdfsubject={scientific publication}
   ,citecolor={blue},
   linkcolor={blue},
   urlcolor={blue},
   filecolor={blue}
}

\newcommand{\eVdist}{\kern-0.06em}
\newcommand{\gev}{\:\text{Ge\eVdist V}}
\newcommand{\tev}{\:\text{Te\eVdist V}}
\newcommand{\s}{\:\text{s}}

\DeclareMathOperator{\re}{Re}

\newcommand{\D}{\mathrm{d}}
\newcommand{\I}{\mathrm{i}}

\allowdisplaybreaks[1]

\begin{document}

\date{\mbox{ }}

\title{ 
{\normalsize     
\today\  \hfill\mbox{} DESY 13-142\\}
\vspace{2cm}
NMSSM with Gravitino Dark Matter\\
 to be tested at LHC
\\[8mm]}
%
%\vspace{2cm} 
\author{Jasper Hasenkamp$^{a,b}$ and Martin Wolfgang Winkler$^c$\\[2mm]
{\small\it a Center for Cosmology and Particle Physics, New York University, New York, NY, USA}\\
{\small\it b II.~Institute for Theoretical Physics, University of Hamburg, Hamburg, Germany}\\
{\small\it c Deutsches Elektronen-Synchrotron DESY, Hamburg, Germany}\\
{\small\tt Jasper.Hasenkamp@nyu.edu, martin.winkler@desy.de}
}
\maketitle

\thispagestyle{empty}

\vspace{1cm}
 \begin{abstract}
 \noindent
 We present a solution to the gravitino problem, which arises in the NMSSM, allowing for sparticle spectra from ordinary gravity-mediated supersymmetry breaking with weak-scale gravitino dark matter.
 The coupling, which links the singlet to the MSSM sector,  enhances the tree-level Higgs mass, providing an attractive explanation why the observed Higgs boson is so heavy.
The same coupling induces very efficient pair-annihilation processes of the neutralino NLSP.
Its relic abundance can be sufficiently suppressed to satisfy the strong constraints on late decaying relics from primordial nucleosynthesis -- even for very long neutralino lifetimes. 
The striking prediction of this scenario is the detection of a pseudoscalar Higgs boson in the search for top-top resonances at LHC-14, rendering it completely testable. 
 \end{abstract}

\newpage

\section{Introduction}

The Minimal Supersymmetric Standard Model (MSSM) is certainly the most popular candidate for physics beyond the Standard Model (SM).
 One of its virtues is a new stable particle which might form the dark matter of the Universe.
However, despite a hunt lasting for decades, dark matter particle searches failed to provide conclusive signals. 
Searches for supersymmetry (SUSY) at the Large Hadron Collider (LHC) continuously strengthen the limits on supersymmetric particles (see e.g.~\cite{Chatrchyan:2013wxa,ATLAS:2013}).
Moreover, the discovery of the Higgs boson at a mass of $126\gev$~\cite{Aad:2012tfa,Chatrchyan:2012ufa} seems to hint at a SUSY spectrum well beyond the TeV scale which threatens parts of its theoretical motivation and would render SUSY unobservable at any collider experiment in the foreseeable future.

Turning to cosmology, supersymmetric theories are plagued by the presence of the gravitino in the particle spectrum~\cite{Weinberg:1982zq}.
We refer to the plethora of connected problems as 'the gravitino problem'.
Confronted with a naturally late-decaying gravitino that would spoil the success of primordial nucleosynthesis~\cite{Khlopov:1984pf,Ellis:1984eq},
many solutions rely on sophisticated schemes of SUSY breaking, non-standard cosmologies without observational motivation and are inconsistent with the most appealing explanations for the origin of matter, in particular, baryogenesis via thermal leptogenesis~\cite{Fukugita:1986hr}.

In this work we present a solution to the gravitino problem free of these shortcomings which arises in the Next-to-Minimal Supersymmetric Standard Model (NMSSM).
The main motivation to study the NMSSM arises from the additional tree-level contribution to the Higgs mass which allows for $m_h=126\gev$ with only moderate loop corrections from the stop sector.
Consequently, the NMSSM can be embedded nicely into a scheme of ordinary gravity-mediated SUSY breaking with all superpartners around the TeV scale and, thus, in the discovery reach of the LHC.
If the gravitino is the lightest supersymmetric particle (LSP), thermally produced gravitinos might form the observed dark matter~\cite{Bolz:2000fu} implying a non-detection in any, direct or indirect, particle dark matter search.
For gravitino masses around the weak scale the required reheating temperature of the universe after inflation is sufficiently high to be compatible with the most attractive mechanisms of baryogenesis like standard thermal leptogenesis.
While in this scenario the gravitino is stable, strong constraints arise from the late decay of the next-to-lightest supersymmetric particle (NLSP) which might spoil the success of primordial nucleosynthesis~\cite{Ellis:1984er,Moroi:1993mb} or lead to distortions in the cosmic microwave background (CMB)~\cite{Hu:1993gc,Fixsen:1996nj}.

We shall point out that the NMSSM allows for very efficient pair-annihilation processes of the lightest neutralino. The freeze-out density of a neutralino NLSP can be sufficiently reduced to satisfy the cosmological bounds. 
Against naive intuition, the lowest neutralino abundances occur for a mixed singlino/higgsino NLSP rather than for a wino.
 The annihilations are driven by the coupling $\lambda$ which links the NMSSM-singlet to the MSSM sector. 
Remarkably, it is the same coupling which enhances the tree-level Higgs mass.

Important implications for Higgs searches at the LHC arise. 
Most remarkably, we find that the NMSSM with gravitino dark matter predicts the detection of a pseudoscalar Higgs in the search for $t\bar{t}$-resonances at LHC-14.

The paper is organized as follows: 
In Sec.~\ref{sec:cosmoconstr}, we compile and elaborate on the cosmological constraints applying to late-decaying neutralinos.
Sec.~\ref{sec:soltogravprob} presents the solution to the gravitino problem in the NMSSM.
The implications for Higgs searches at the LHC are provided in Sec.~\ref{sec:HiggsLHC}, before we conclude in the last section.

\section{Cosmological constraints on late-decaying neutralinos}
\label{sec:cosmoconstr}
In this section we compile and elaborate on cosmological constraints applying to the neutralino NLSP. Depending on the neutralino lifetime these will either arise from the requirement of successful BBN or observations of the cosmic microwave background (CMB).
We will focus on the case where both particles, NLSP and LSP, have masses around the weak scale, $m=\mathcal{O}(100\gev)$, as it is expected from gravity-mediated SUSY breaking. Nevertheless, our constraints can be translated to other scenarios and decaying particles.
\subsection{CMB and BBN constraints}
As long as $Z$-bosons in the final state are kinematically inaccessible, the lightest neutralino almost always decays into gravitino and photon\footnote{This statement holds unless the photino component of the lightest neutralino is strongly suppressed and unless the superpartner of another neutralino is lighter than the $Z$-boson. The latter case only appears in extensions of the MSSM, e.g. in the NMSSM a singlino NLSP could decay into a gravitino and a light singlet-like Higgs.}. The corresponding decay rate reads~\cite{Feng:2004mt}
\begin{equation}\label{eq:decayrate}
\Gamma_\chi (\widetilde{\chi}_1 \rightarrow \Psi_{3/2} + \gamma) ~=~ \frac{|\text{N}_{\widetilde{\gamma}}|^2\:m_{\chi}^5 }{48 \pi M_P^2 \: m_{3/2}^2} \,\left(1-\left(\frac{m_{3/2}}{m_{\chi}}\right)^2\right)^3 \left(1 + 3\,\left(\frac{m_{3/2}}{m_{\chi}}\right)^2\right)\; ,
\end{equation}
where $|\text{N}_{\widetilde{\gamma}}|^2$ and $m_{\chi}$ denote the photino fraction and the mass of the lightest neutralino, respectively. In case $m_{\chi}-m_{3/2} \ll m_{\chi}$, the decay rate can be approximated as
\begin{equation}\label{eq:decayratedegnerate}
\Gamma_\chi (\widetilde{\chi}_1 \rightarrow \Psi_{3/2} + \gamma) ~\simeq~ \frac{2\,|\text{N}_{\widetilde{\gamma}}|^2 }{3 \pi M_P^2 }\:(m_{\chi}-m_{3/2})^3 \; .
\end{equation}
The decay is suppressed by the Planck mass $M_P$, which makes the neutralino NLSP extremely long-lived with typical lifetimes $\tau=\Gamma_\chi^{-1} = 10^7-10^{12}\s$. We now turn to the cosmological bounds on the emission of energetic photons, where we concentrate on this time window. The constraints depend on the overall amount of electromagnetic energy injected into the plasma, more specifically on the product $\Omega_\chi h^2 \times \Delta_\text{em}$. Here $\Omega_\chi h^2$ denotes the relic density of the neutralino in units of today's critical energy density, if it had not decayed. We call this quantity the freeze-out density in the following. The fraction of neutralino mass converted into electromagnetic energy per decay is given as the product of the branching ratio $\text{Br}(\widetilde{\chi}_1 \rightarrow \Psi_{3/2} + \gamma)$ and the energy fraction carried by the photon which is determined by two-body kinematics. We find 
\begin{equation}
\label{eq:emenergy}
 \Delta_\text{em}~=~ \frac{m_\chi^2-m_{3/2}^2}{2\,m_\chi^2} \;\text{Br}(\widetilde{\chi}_1 \rightarrow \Psi_{3/2} + \gamma)\;.
\end{equation}
Below the $Z$-boson threshold, we expect $\text{Br}(\widetilde{\chi}_1 \rightarrow \Psi_{3/2} + \gamma)\simeq 1$.

\begin{figure}[t]
\begin{center}
\includegraphics[width=9cm]{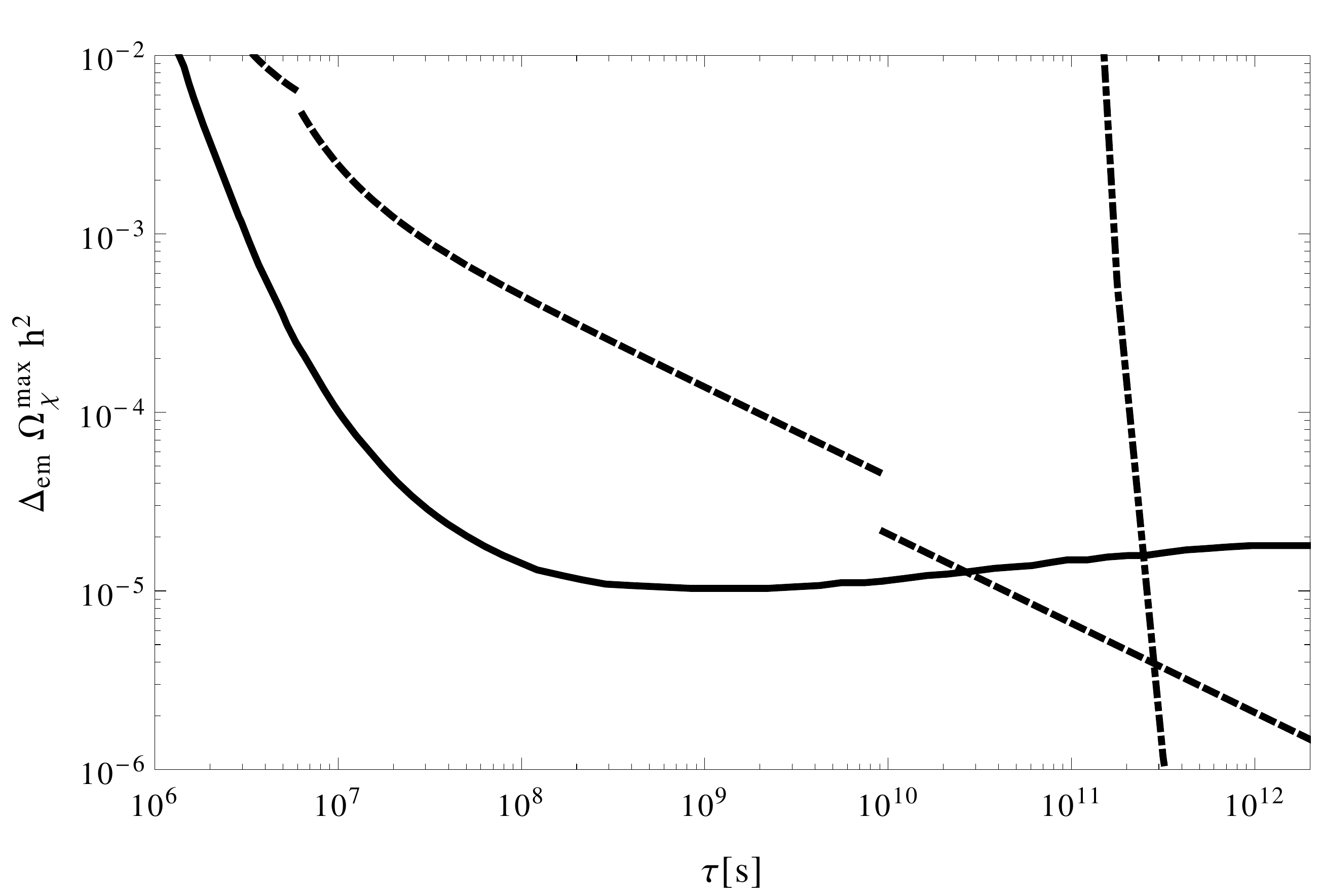}
\end{center}
\caption[BBN+CMB constraints]{\footnotesize{
Cosmological constraints on the injected electromagnetically-interacting energy density $\Omega_\chi^\text{max} h^2 \times \Delta_\text{em}$ depending on the time of decay $\tau$ as arising from BBN (solid), CMB spectral distortions (dashed) and the ionization history of the universe as observable in the CMB (dash-dotted).
The area above the curves is considered as excluded.
}}
\label{fig:constraints}
\end{figure}

\paragraph*{Primordial nucleosynthesis}
During BBN light nuclei like Helium are synthesized in the primordial plasma. Occurring at cosmic times between $1\s$ and $10^3\s$ BBN provides the deepest insight into the early universe. The predicted abundances are in convincing agreement with observations today~\cite{Beringer:1900zz}.
An energetic particle injected into the plasma even after BBN can disrupt the formerly build nuclei, spoiling the success of BBN. Therefore, the requirement of successful BBN leads to stringent bounds on the freeze-out energy density of late-decaying particles even if they decay long after the synthesis has ended.  We adopt bounds determined in~\cite{Jedamzik:2006xz} (see references therein and~\cite{Kawasaki:2004qu} for earlier work on BBN bounds). 
In Fig.~\ref{fig:constraints} the bound on the electromagnetically interacting energy density injected into the plasma  
is depicted as solid line.

\paragraph*{Cosmic microwave background}
 For the lifetimes under consideration the emission of particles with SM interactions leads to spectral distortions of the CMB, because the photons do no longer thermalize completely.
Updating the analysis of~\cite{Hu:1992dc} by taking into account the current limits, obtained by COBE FIRAS, on the chemical potential $|\mu|<9 \times 10^{-5} = \mu_\text{max}$ and the Compton-$y$ parameter $y<1.5 \times 10^{-5}=y_\text{max}$~\cite{Mather:1993ij,Fixsen:1996nj} we find
\begin{equation}
\label{specdisbound}
\Omega_\chi^\text{max} \times \Delta_\text{em} ~=~ 0.94 \, \mu_\text{max}  \left(\frac{10^{10} \text{ s}}{\tau}\right)^\frac{1}{2} 
e^{\left(t_\mu/\tau \right)^{5/4}} \, ,
\end{equation}
where $t_\mu \simeq 6.91 \times 10^6\s\;(1-Y_p/2)^{4/5} = 6.2\times 10^6\s$ denotes the timescale of thermalization in the universe at that epoch with its largest uncertainty inherited from the  error in the primordial Helium abundance $Y_p \simeq 0.249$. We find that this error is completely negligible and insert PDG mean values~\cite{Beringer:1900zz} throughout.
In Fig.~\ref{fig:constraints} the corresponding bound is depicted  as dashed line.

For times earlier than $t_\mu$ we replace $e^{(t_\mu/\tau)^{5/4}} \rightarrow 0.48 (\tau/t_\mu)^{10/18} e^{1.99(t_\mu/\tau)^{10/18}}$ following~\cite{Chluba:2011hw}. 
 For times $\tau \lesssim 9 \times 10^{9}$ s  a Bose-Einstein spectrum is established by elastic Compton scattering. For any point in Fig.~\ref{fig:constraints} the number of injected photons is negligible relative to the number of background photons and their energy density is small compared to the energy density of background photons $\rho_\gamma$. Then the induced chemical potential is proportional to the injected energy density, $ 1.40 \mu = \, \Delta_\text{em} \rho_\chi/\rho_\gamma$. For later times the spectrum can be described by the Compton-$y$ parameter, which is induced as $\Delta_\text{em} \rho_\chi/\rho_\gamma = 4 y$. 
In this regime one has to replace $\mu_\text{max}\rightarrow (4/1.4) \,y_\text{max}$ in~\eqref{specdisbound}. 
The jump in the bound due to this change in the description of the spectrum with the corresponding constraints is easily identified at the corresponding time.
At $2.6 \times 10^{10}$ s the bound from CMB spectral distortions becomes tighter than the BBN bound.  

The injection of particles with SM interactions may change the ionisation history of the Universe, which can leave observable consequences in the CMB (see e.g.~\cite{Slatyer:2012yq}). Besides the dependence on $\tau$, bounds differ for different decay products~\cite{Cline:2013fm}. In Fig.~\ref{fig:constraints} we depict the bound applying to our case, i.e.\@ photons emitted with energies of some tens of GeV, as dash-dotted line.\footnote{We are thankful to Jim Cline who provided this curve extending the computation in~\cite{Cline:2013fm} for the corresponding bound towards the earlier times we are interested, here. We find that a quadratic extrapolation approximates the true bound fairly well and can be used instead, while linear and cubic extrapolations deviate significantly from the true bound.}
We see that it serves as a sharp cut at a lifetime $\tau = 3 \times 10^{11}$ s. 

\subsection{Constraint on the neutralino freeze-out density}

\begin{figure}[t]
\begin{center}
\includegraphics[width=7cm]{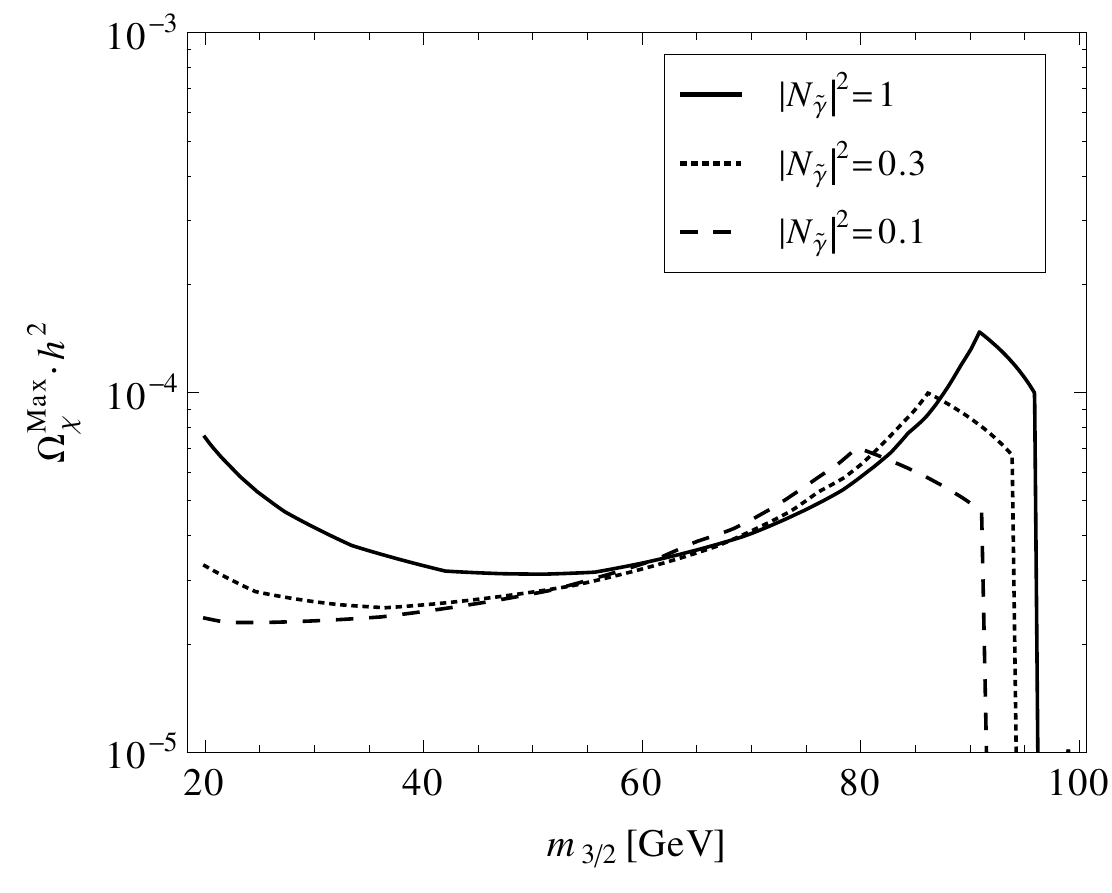}
\end{center}
\caption{\footnotesize{Constraints on the freeze-out density of a neutralino NLSP for different values of the photino fraction $|\text{N}_{\widetilde{\gamma}}|^2$. The neutralino mass is set to $m_{\chi}=100\gev$.}}
\label{fig:neutralinoconstraints}
\end{figure}

In Fig.~\ref{fig:neutralinoconstraints} we depict the resulting BBN + CMB constraints on the neutralino freeze-out density for different choices of the photino fraction. For the gravitino masses depicted, the neutralino lifetime is $\tau>10^7\s$. In this regime the BBN constraint on $\Omega_{\chi} h^2 \times \Delta_\text{em}$ remains almost constant. As the amount of energy carried away by the gravitino grows with increasing $m_{3/2}$, the bound on $\Omega_{\chi} h^2$ becomes weaker for larger $m_{3/2}$. But for very long neutralino lifetimes, the CMB constraints jump in, leading to the sharp decrease at $m_{3/2}\sim 0.8\dots 0.9\, m_{\chi}$.\footnote{Only for very tiny mass splittings $m_{\chi}-m_{3/2}=\mathcal{O}(\text{MeV})$, the constraint on $\Omega_\chi h^2$ becomes weaker again~\cite{Boubekeur:2010nt}.}

The weakest bound on $\Omega_\chi h^2$ -- corresponding to the maxima in Fig.~\ref{fig:neutralinoconstraints} -- is obtained for the lifetime $\Gamma_\chi^{-1} = 2.6\times 10^{10}\s$ where the bounds from BBN and CMB spectral distortions intersect. The gravitino mass corresponding to this lifetime can be obtained by inverting~\eqref{eq:decayratedegnerate}
\begin{equation}
  m_{3/2} ~\simeq~ m_{\chi} - \frac{8.9\gev}{|\text{N}_{\widetilde{\gamma}}|^{2/3}}\;.
\end{equation}
By use of~\eqref{eq:emenergy} and the bound from Fig.~\ref{fig:constraints}, we find that the corresponding limit on the freeze-out density reads
\begin{equation}\label{eq:limitomega}
 \Omega_{\chi} h^2~ <~ 1.4\cdot 10^{-4}\;|\text{N}_{\widetilde{\gamma}}|^{2/3}\,\left(\frac{m_{\chi}}{100\gev}\right)\;.
\end{equation}
If one imposes this requirement, one can be sure that the neutralino NLSP does not spoil cosmological observations at least for some range of weak scale gravitino masses.

We have concentrated so far on the case where only the electromagnetic decay of the lightest neutralino is kinematically accessible. One might speculate that the constraints can be relaxed if further decay modes like $\widetilde{\chi}_1 \rightarrow \Psi_{3/2} + Z$ open up. In this case, a significant energy fraction would be induced in form of quark pairs arising from $Z$ boson decay. However, the emission of hadronically interacting particles always induces the emission of electromagnetically interacting secondaries like photons or electrons, for example, from pion decay. Therefore, the bounds on hadronically interacting and electromagnetically interacting decay products are very similar for $\tau>10^7\s$. For lifetimes $\tau< 10^7\s$, the electromagnetic bounds get significantly weaker. But in order to realize such short lifetimes a large mass splitting $m_{\chi}-m_{3/2}$ is required and, in this case, the hadronic branching ratio is always sizeable. Other than the electromagnetic constraints, the hadronic BBN constraints are very tight also for NLSP lifetimes $\tau< 10^7\s$.
In order to evade the hadronic constraints, $\tau < 10^4\s$ is required corresponding to $m_{3/2}=\mathcal{O}(\text{GeV})$~\cite{Covi:2009bk}.\footnote{Another not very appealing possibility to realize $\tau < 10^4\s$ consists in putting $m_\chi$ and $m_{3/2}$ above the TeV scale.} Such low gravitino masses are incompatible with gravity-mediated SUSY breaking and, in addition, would reintroduce the gravitino problem through thermal gravitino overproduction. They shall not be considered here. 

For gravitino masses as expected in gravity mediation, the cosmological constraints can thus not be considerably relaxed even if further decay modes of the neutralino open up. The weakest constraint on $\Omega_{\chi} h^2$ is obtained if the neutralino and gravitino are close in mass, but still sufficiently apart to avoid the CMB constraints, as described above. This situation lead to~\eqref{eq:limitomega} which we will take as the relevant constraint in the following.

\subsection{Implications for the annihilation cross section}\label{sec:limitannihilation}

The freeze-out density of the neutralino NLSP can be obtained by solving the corresponding Boltzmann equations. For this purpose -- as the neutralino decays long after its decoupling from thermal equilibrium -- it can be treated as a stable particle. We find
\begin{equation}\label{eq:relicdensity}
 \Omega_\chi h^2~ \simeq~ 9.3\cdot 10^{-12}\gev^{-2} \times \frac{m_{\chi}/\mathrm{T}_\mathrm{F}}{\widetilde{\sigma }}\;,
\end{equation}
where we have defined
\begin{equation}
 \widetilde{\sigma}~ =~\frac{1}{\mathrm{T}_\mathrm{F}}\int\limits_0^{\mathrm{T}_\mathrm{F}} \D\text{T} \;\langle \sigma \mathscr{\text{v}}\rangle
\end{equation}
with $\langle\sigma \mathscr{\text{v}} \rangle$ denoting the thermally averaged annihilation cross section and $\mathrm{T}_\mathrm{F}$ the freeze-out temperature. 
If $\langle \sigma \mathscr{\text{v}}\rangle$ is temperature-independent, we can replace $\widetilde{\sigma }$ by $\langle \sigma \mathscr{\text{v}}\rangle$ in~\eqref{eq:relicdensity} as frequently done in the literature.

Taking into account that for $\Omega_\chi h^2=\mathcal{O}(10^{-4})$, the freeze-out temperature can be estimated as $\mathrm{T}_\mathrm{F}$ = $m_{\chi}/30$, we can directly translate the constraint~\eqref{eq:limitomega} into a limit on the annihilation cross section
\begin{equation}\label{eq:mincross}
\widetilde{\sigma } ~>~ \frac{1}{16\pi\, m_{\chi}^2} \; \left(\frac{m_{\chi}}{100\gev}\right)\frac{1}{|\text{N}_{\widetilde{\gamma}}|^{2/3}}\;.
\end{equation}
We would like to recall that the dominant velocity-independent contribution to the cross section can be written in the form $ \widetilde{\sigma } = \frac{g_\text{eff}^2\, \mathcal{F}}{16\pi\, m_{\chi}^2}$, where $g_\text{eff}$ denotes an effective coupling, while $\mathcal{F}$ is a dimensionless function of the relevant particle masses -- for massless intermediate and final states $\mathcal{F}=1$. Therefore,~\eqref{eq:mincross} can only be satisfied for very large effective couplings $g_\text{eff}>1$ and/ or for $\mathcal{F}>1$ which in turn requires resonant enhancement.

\section{A solution to the gravitino problem in the NMSSM}
\label{sec:soltogravprob}
Since the discovery of the Higgs boson at the LHC, singlet extensions of the MSSM have received particular attention due to their ability to increase the tree-level Higgs mass beyond $M_Z$. In order to affect the Higgs mass considerably, a sufficiently large coupling $\lambda$ between the MSSM Higgs fields and the singlet is required. We shall now discuss the relevant annihilation processes of a long-lived neutralino NLSP which are driven by this coupling. In particular, we are going to identify regions in the NMSSM parameter space which are consistent with weak scale gravitino dark matter, i.e.\ which satisfy the constraint~\eqref{eq:limitomega}. 

Scenarios to obtain a suppressed freeze-out density of a general NLSP have been discussed in the context of the MSSM. These include the case of a neutralino NLSP with resonant annihilation~\cite{Covi:2009bk} or coannihilations~\cite{Covi:2010au}, a stau NLSP with an enhanced coupling to Higgs bosons~\cite{Ratz:2008qh,Pradler:2008qc} and a stop NLSP~\cite{Berger:2008ti,Kusakabe:2009jt}. Alternatively, the NLSP density can be diluted by entropy production after its freeze-out~\cite{Hasenkamp:2010if}, implying a non-standard expansion history of the universe observable in the primordial gravitational wave background~\cite{Durrer:2011bi}.
While these possibilities also exist within the NMSSM, we will focus here on the NMSSM specific solution to the gravitino problem.

\subsection{Neutralinos and Higgs bosons in the NMSSM}\label{sec:notation}

We consider the simplest $Z_3$-symmetric version of the NMSSM which is defined by the superpotential
\begin{equation}
 W ~=~ \lambda S H_u H_d + \frac{\kappa}{3}\,S^3\;.
\end{equation}
The $Z_3$ should only be an approximate symmetry, otherwise its breaking at the electroweak scale would induce disastrous domain walls~\cite{Vilenkin:1984ib}. But we restrict our attention to the case, where the $Z_3$-breaking effects are sufficiently suppressed and do not play a role for phenomenology (see discussion in~\cite{Ellwanger:2009dp}).
The soft terms corresponding to the superpotential above read
\begin{equation}
 V_{\text{soft}} ~=~ m_{h_u} |h_u|^2+ m_{h_d} |h_d|^2+ m_{s} |s|^2 + (\lambda A_\lambda h_u  h_d + \frac{\kappa}{3} A_\kappa s^3 +\text{h.c.})\;.
\end{equation}
We have denoted superfields and their scalar components by capital and small letters, respectively. The lightest neutralino, which is the NLSP in our scenario, is a linear combination of the singlet fermion (singlino) $\widetilde{s}$ as well as the higgsinos $\widetilde{h}_{d,u}$ and gauginos $\widetilde{B},\widetilde{W}$:
\begin{equation}
 \widetilde{\chi}_1 ~=~ \text{N}_{11}\, (-\I\, \widetilde{B}) + \text{N}_{12}\, (-\I\, \widetilde{W}) + \text{N}_{13}\, \widetilde{h}_d+ \text{N}_{14}\, \widetilde{h}_u+ \text{N}_{15}\,\widetilde{s}\;.
\end{equation}
The coefficients $\text{N}_{1i}$ can be determined by diagonalizing the neutralino mass matrix which can be found in Appendix~\ref{sec:neutralinomatrix}. The photino component, which is important for the neutralino lifetime, is given as 
\begin{equation}
\text{N}_{\widetilde{\gamma}}~=~\text{N}_{11} \cos\theta_W + \text{N}_{12} \sin\theta_W 
\end{equation}
with $\theta_W$ denoting the Weinberg angle.

Turning to the Higgs sector, we use the basis $(a,a_s)$ for the pseudoscalar Higgs fields, where $a$ and $a_s$ denote the MSSM and singlet pseudoscalar, respectively. The corresponding mass matrix is given in Appendix~\ref{sec:higgsmatrix}. For the CP even Higgs bosons, it is instructive to define the basis $(H,h,h_s)$ via
\begin{equation}
 \begin{pmatrix}  H\\  h\\  h_s \end{pmatrix}
~=~\sqrt{2}\,\begin{pmatrix}  \sin\beta & \cos\beta & 0\\ -\cos\beta & \sin\beta & 0\\  0 & 0 & 1 \end{pmatrix}\, 
\begin{pmatrix} \re (h_d^0) - v_d\\  \re (h_u^0)-v_u\\  \re (s)-v_s \end{pmatrix}\;.
\end{equation}
Here $v_{u,d,s}$ denote the vacuum expectation values of the Higgs fields with $\sqrt{v_u^2+v_d^2}\simeq 174\gev$ and $\tan\beta= v_u/v_d$. In this basis the Higgs couplings have a simple form: $h$ couples to gauge bosons and fermions exactly as the SM Higgs, while $H$ does not couple to gauge bosons and its couplings to down-type (up-type) fermions include a factor $\tan\beta$ ($\cot\beta$) compared to the SM Higgs~\cite{Miller:2003ay}. The mass matrix in this basis can be found in Appendix~\ref{sec:higgsmatrix}.
The mass eigenstates, obtained from diagonalizing the Higgs mass matrices, shall be denoted by $h_{1,2,3}$ and $a_{1,2}$ ordered as usual by their mass.

\subsection{Constraint from perturbativity}

It has been pointed out that there arises an upper bound on the couplings $\lambda$ and $\kappa$ if one requires the NMSSM to remain perturbative up to the GUT scale~\cite{Haber:1986gz,Espinosa:1991gr}.  Neglecting a slight dependence on the superpartner spectrum\footnote{We have obtained the constraint with NMSSMTools~3.2.3~\cite{Ellwanger:2004xm,Ellwanger:2006rn,Das:2011dg}, where we set the superpartner masses to $2\tev$ and the top quark mass to $m_t=172\gev$.}, we approximate
\begin{equation}\label{eq:perturbativitybound}
 \lambda^2 + \kappa^2 ~<~ 0.6\, (1 - \tan\beta^{-3})\;.
\end{equation}
The above estimate holds to the accuracy of a few per cent for $\tan\beta \gtrsim 1.3$. This is sufficient for our purposes as thresholds effects and higher order corrections are expected to affect the constraint at this level of precision. At $\tan\beta\simeq 1.3$ the top Yukawa coupling $y_t$ becomes non-perturbative below the GUT scale independent of $\lambda$ and $\kappa$. Therefore we restrict our attention to $\tan\beta > 1.3$.

Several possibilities to circumvent the perturbativity constraint have been discussed in the literature~\cite{Masip:1998jc,Harnik:2003rs,Chang:2004db,Hardy:2012ef}. In particular, the RGE running of $\lambda$ can be affected by the inclusion of extra matter below the GUT scale. To keep the model minimal we will, nevertheless, apply~\eqref{eq:perturbativitybound} and -- if relevant -- point out the consequences of dropping this constraint.

\subsection{Efficient neutralino annihilation in the NMSSM}\label{sec:channel}

Let us now turn to the relevant annihilation processes of the lightest neutralino in the NMSSM. In order to be consistent with gravitino dark matter, the neutralino freeze-out density must be suppressed by very efficient pair-annihilation processes. Indeed, by inspecting~\eqref{eq:mincross}, one can already deduce that couplings larger than unity are required to satisfy the cosmological bound -- unless in the case of a resonance.

In the MSSM, even a wino does not have a sufficiently large annihilation cross section to be consistent with~\eqref{eq:limitomega}~\cite{Covi:2009bk}. While in the NMSSM, $\lambda$ can be somewhat larger than $g_2$, the perturbativity bound prevents us from obtaining $\lambda>1$. If we restrict $\lambda$ to the allowed range, some resonant enhancement is still required. 

\begin{figure}[t]
\begin{center}
\includegraphics[width=4.5cm]{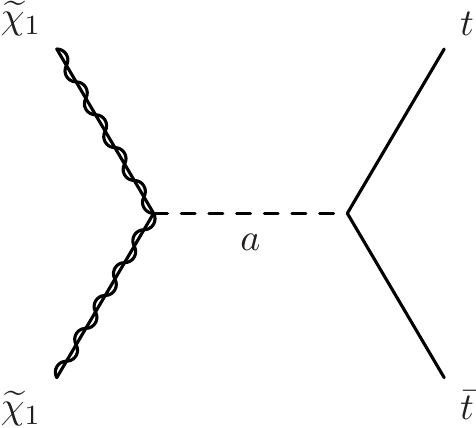}
\end{center}
\caption{\footnotesize{Neutralino annihilation into top quarks via an intermediate pseudoscalar Higgs.}}
\label{fig:feynman}
\end{figure}

We focus on processes involving a pseudoscalar in the s-channel as annihilation via CP even scalars is velocity-suppressed. The largest cross section can be obtained for the process $ \widetilde{\chi}_1  \widetilde{\chi}_1 \rightarrow a^* \rightarrow \bar{t}t$ with a MSSM like pseudoscalar in the intermediate state. The corresponding Feynman diagram is shown in Fig.~\ref{fig:feynman}. 

In our parameter scans which follow later on, we did not impose this annihilation channel to dominate. Nevertheless, we find that whenever a sufficiently low neutralino freeze-out density is obtained, it is virtually always induced by the process above. This implies that the lightest neutralino is very likely to be heavier than the top quark in this scenario. In principle, the trilinear soft term $\lambda A_\lambda$ could also drive efficient annihilation into Higgs or $Z$ bosons.\footnote{Note that the $Z$ boson contains the Goldstone mode which couples as a pseudoscalar Higgs.} But a too large $A_\lambda$  is excluded as it would decrease the mass of the SM-like Higgs via singlet-doublet mixing. In addition, it would lead to a decoupling of the MSSM-like pseudoscalar. Mainly below the top threshold, we still find corners in the NMSSM parameter space with $\Omega_\chi h^2 < 10^{-4}$ via annihilation into Higgs / $Z$ bosons. But we will now concentrate on the case with top quarks in the final state which appears much more frequently. 

In order to determine the neutralino freeze-out density we use NMSSMTools~3.2.3~\cite{Ellwanger:2004xm,Ellwanger:2006rn,Das:2011dg} interfaced with micrOMEGAs~2.2~\cite{Belanger:2008sj} which performs a full numerical solution to the Boltzmann equation. Still, it is instructive to gain also some analytic understanding of the dependence of $\Omega_\chi h^2$  on the NMSSM parameters. Closed expressions for the cross section $\widetilde{\sigma}$ in the vicinity of the pseudoscalar resonance can only be obtained under simplifying assumptions: one has to neglect either the width of the resonance (narrow width approximation) or the kinetic energy of the WIMPs (broad width approximation). Unfortunately, as the width of the MSSM pseudoscalar is neither very large nor very small, none of the two limits is strictly applicable. Still, we find that a reasonable estimate can be obtained by using the broad width approximation for $m_{a}< 2\,m_{\chi}$ and the narrow width approximation in the opposite regime. Using this approach, we obtain
\begin{equation}\label{eq:approximation}
 \widetilde{\sigma} ~\simeq~
\begin{cases}
 \frac{3\,g^2_{a\chi\chi} \,y_t^2\cos^2{\beta}}{4\pi }\,
 \frac{m_{\chi}\,\sqrt{m_{\chi}^2-m_t^2}}{(4 m_{\chi}^2 -m_a^2)^2+m_a^2 \,\Gamma_a^2} \quad\; & \text{for}\quad m_{a_2}< 2\,m_{\chi}   \;,       \\[4mm]
 \frac{ g^2_{a\chi\chi} }{m_a}\frac{\pi}{\mathrm{T}_{\mathrm{F}}}\;\,\text{erfc}\sqrt{\frac{1}{\mathrm{T}_{\mathrm{F}}}\frac{m_a^2- 4 m_{\chi}^2}{4 m_{\chi}}} 
 \quad\; & \text{for}\quad m_{a_2}> 2\,m_{\chi}      \;,                     
\end{cases}
\end{equation}
where we have neglected the mixing of the MSSM pseudoscalar with the singlet. Here erfc denotes the complementary error function and $\Gamma_a$ the pseudoscalar decay width which is given as
\begin{equation}
\Gamma_a ~=~ \frac{3\,y_t^2\cos^2{\beta}}{16\,\pi}\,\sqrt{m_a^2-4 m_t^2}\;,
\end{equation}
where we considered only the dominant decay mode $a \rightarrow \bar{t}t$. The coupling between the lightest neutralino and the MSSM pseudoscalar reads~\cite{Ellwanger:2009dp}
\begin{eqnarray}
 g_{a\chi\chi} &=&  g_2 \,( \text{N}_{12} \text{N}_{13}\, \sin\beta-  \text{N}_{12} \text{N}_{14}\,\cos\beta)
- g_1\, (\text{N}_{11} \text{N}_{13}\,\sin\beta  -  \text{N}_{11} \text{N}_{14}\,\cos\beta)\nonumber\\ &&  + \sqrt{2} \,\lambda \,( \text{N}_{14} \text{N}_{15}\,\sin\beta +  \text{N}_{13} \text{N}_{15}\,\cos\beta)\;.
\end{eqnarray}
We would like to point out that in the NMSSM this coupling can be considerably stronger than in the MSSM. This is due to the appearance of an additional factor $\sqrt{2}$ in front of $\lambda$ and due to the fact that $\lambda$ can be larger than the gauge couplings $g_1$ and $g_2$. In gravity mediation, one expects the bino to be lighter than the wino due to gaugino mass unification at the high scale, which further suppresses $g_{a\chi\chi}$ in the MSSM. If we require gaugino mass unification, the maximal $g_{a\chi\chi}$ in the NMSSM is a factor $\sqrt{2}\lambda / g_1$ larger than in the MSSM.\footnote{Note that in the NMSSM and MSSM the coupling $g_{a\chi\chi}$ is maximized in different regions of the parameter space. In the MSSM, large bino-higgsino mixing is required, while in the NMSSM large singlino-higgsino mixing is required.} For $\lambda\simeq 0.7$ this corresponds to a factor $\sim 3$ in the coupling which translates into an enhancement of the annihilation cross section by almost one order of magnitude. In this sense, the situation in the NMSSM is more favorable than in the MSSM as one needs considerably less tuning of $m_a$ in order to suppress the freeze-out density. 

\begin{figure}[t]
\begin{center}
\includegraphics[width=9cm]{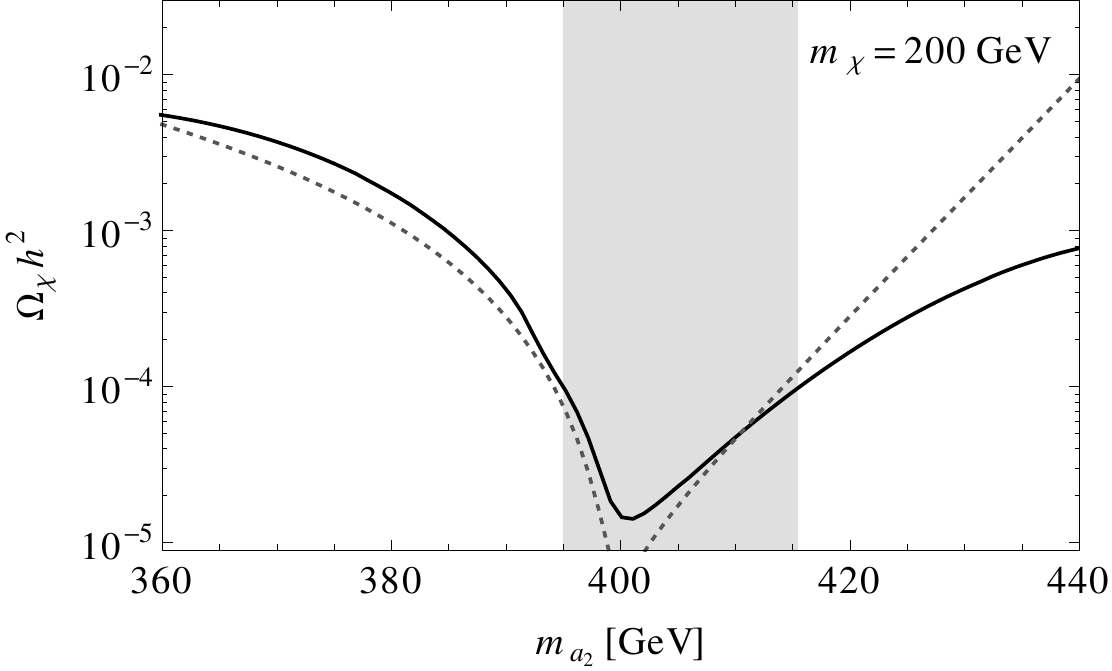}
\end{center}
\caption{\footnotesize{Dependence of the neutralino freeze-out density $\Omega_\chi h^2$ on the mass of the MSSM-like pseudoscalar $a_2$. The solid line shows the numerical result obtained with micrOMEGAs, the dashed contours refer to the analytical approximation~\eqref{eq:approximation}. To guide the eye, the range of $m_{a_2}$ giving rise to $\Omega_\chi h^2 < 10^{-4}$ is indicated by the shaded band. We have set $\tan\beta=2$, $\lambda=0.65$, $\kappa=0.3$ and assumed sfermion and gaugino masses of $1\tev$.}}
\label{fig:resonance}
\end{figure}

In Fig.~\ref{fig:resonance}, we depict the neutralino freeze-out density as a function of $m_{a_2}$, where $a_2$ is the MSSM-like pseudoscalar in this case. We have chosen $\lambda\simeq 2\kappa$ as the coupling $g_{a\chi\chi}$ is driven by singlino-higgsino mixing. It is thus preferential to have $\widetilde{h}$ and $\widetilde{s}$ at a similar mass in order to maximize the product $\text{N}_{14} \text{N}_{15}$.

It can be seen that $\Omega_\chi h^2 = \mathcal{O}(10^{-4})$ can indeed be realized with mild tuning of $m_{a_2}$ at the level of $10\%$. In the vicinity of the resonance, the analytic formulas presented above agree qualitatively with the numerical result obtained with micrOMEGAs. The narrow width approximation, however, already breaks down for $m_a$ more than $\sim 5\%$ away from the pole.

\subsection{Gravitino problem vs Higgs mass}\label{sec:limittan}

In the last section we have described a mechanism to suppress the neutralino freeze-out density so strongly that its late decay into the gravitino does not spoil cosmological observations. The virtue of this scenario is that the coupling $\lambda$, which is responsible for the efficient neutralino annihilation, at the same time induces a new tree-level contribution to the mass of the SM-like Higgs. 

We wish to recall that the singlet-induced contribution to the Higgs mass is not always positive. There may arise large off-diagonal entries in the Higgs mass matrix, leading to a reduction of $m_{h_1}$ by mixing effects. The Higgs mass matrix in the basis $(H,h,h_s)$ is given in Appendix~\ref{sec:higgsmatrix}. If we apply the MSSM decoupling limit\footnote{The MSSM decoupling limit can be applied as neutralino annihilation into top quarks in the vicinity of the pseudoscalar resonance requires $m_a \gtrsim 2\,m_t$.} we only have to consider the submatrix spanned by $h$ and $h_s$. At tree-level, the mass of the lighter eigenstate, which we take to be the SM-like Higgs, can be approximated as
\begin{equation}\label{eq:higgs1mass}
 m_{h_1}^2 ~\simeq~ m_h^2 - \frac{m_{hh_s}^4}{m_{h_s}^2-m_h^2} ~\simeq~ M_Z^2 \cos^2{2\beta} + \lambda^2 v^2 \sin^2{2\beta} - \frac{\lambda^4 v^2}{\kappa^2}
\left(1 - \frac{m_a^2\sin^2{2\beta}}{4\,\mu^2}\right)^2\, 
\end{equation}
where we neglected terms which are subleading in the parameter region we are interested in. The second term on the right-hand side arises due to the enhanced quartic Higgs coupling in the NMSSM. The last term accounts for the reduction of $m_{h_1}$ by mixing. While one could expect that a large singlet self-coupling $\kappa$ would efficiently reduce the mixing, the combination $\lambda^2 + \kappa^2$ is constrained by the perturbativity bound~\eqref{eq:perturbativitybound}. The more one increases $\kappa$, the smaller $\lambda$ has to be and the smaller becomes the positive contribution $\lambda^2 v^2 \sin^2{2\beta}$ to the Higgs mass. 

As in the MSSM, the Higgs mass receives additional loop contributions mainly from the top/stop sector. To be better off than in the MSSM, we should at least require that
$m_{h_1} > 95\gev$ at tree-level. Otherwise, we would loose the main motivation to study the NMSSM. At the same time, if we want to solve the gravitino problem in the NMSSM, this sets an additional constraint. In order to realize a sufficiently low neutralino freeze-out density, annihilation in the vicinity of the pseudoscalar resonance is required, i.e. $m_a/2 \sim  m_{\chi}< |\mu|$. This implies that
\begin{equation}
 \left(1 - \frac{m_a^2\sin^2{2\beta}}{4\,\mu^2}\right)^2 > \cos^4{2\beta}\;.
\end{equation}
If we additionally impose the perturbativity bound~\eqref{eq:perturbativitybound}, then a tree-level Higgs mass $m_{h_1} > 95\gev$ requires
\begin{equation}
 \tan\beta ~<~ 1.9\;.
\end{equation}
As the production cross sections of the (heavier) Higgs bosons are very sensitive to $\tan\beta$ this will have important consequences for Higgs searches at the LHC.

\begin{figure}[t]
\begin{center}
\includegraphics[width=13cm]{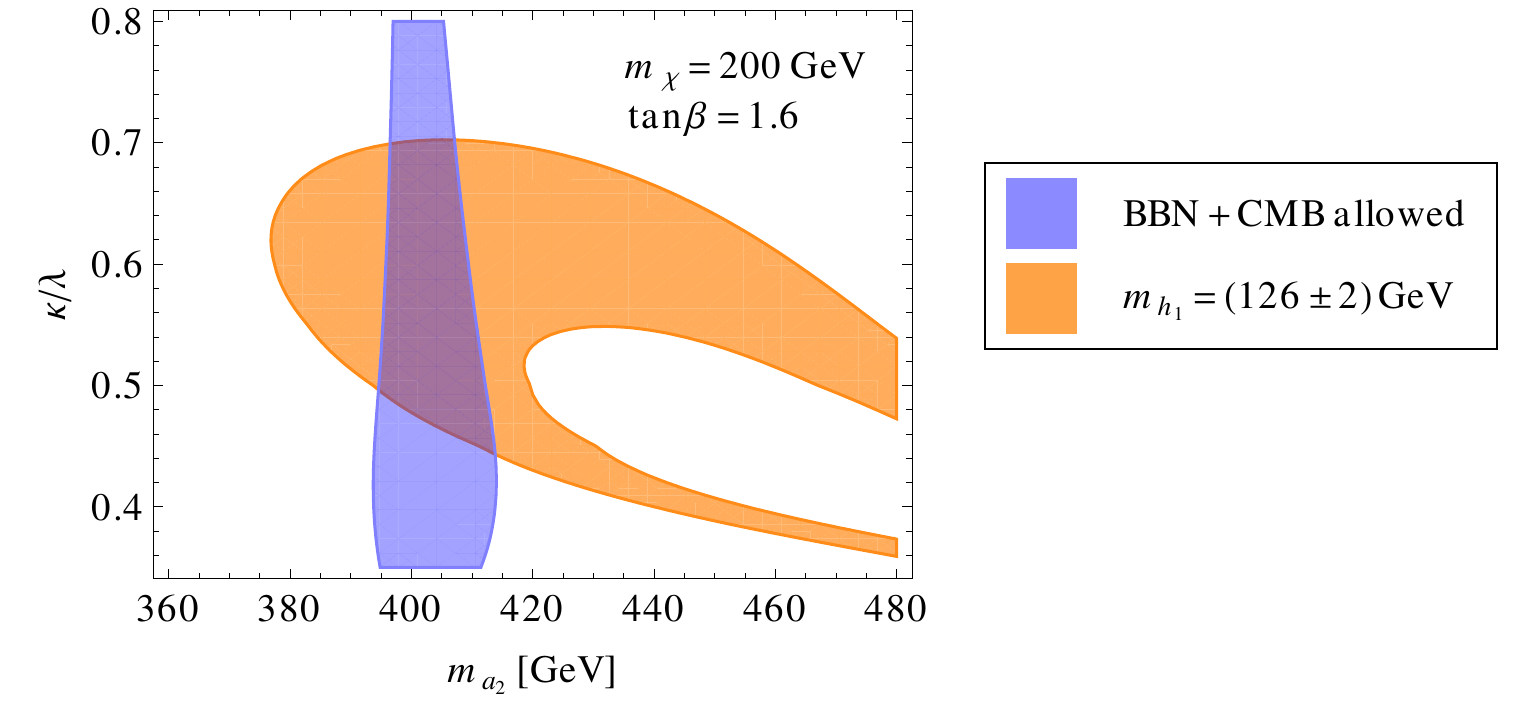}
\end{center}
\caption{\footnotesize{Parameter space in the NMSSM consistent with a gravitino LSP (blue) and a Higgs mass of $m_{h_1}=126 \pm 2\gev$ (orange). The coupling $\lambda$ is set to its maximum allowed by the perturbativity bound~\eqref{eq:perturbativitybound} with the ratio $\kappa/\lambda$ as shown on the y-axis. For each value of the couplings, the effective $\mu$-term is adjusted such that $m_{\chi}$ remains fixed at $200\gev$. The sfermion masses are set to $m_0=2\tev$, the stop trilinear couplings to $A_t=-2\tev$ and the bino mass to $M_1=350\gev$. The remaining gaugino masses are fixed by assuming gaugino mass unification at the high scale.}}
\label{fig:regions}
\end{figure}

In Fig.~\ref{fig:regions}, we provide a parameter scan in the NMSSM which shows the regions consistent with weak scale gravitino dark matter, i.e.\ with the bound~\eqref{eq:limitomega}, as well as with the observed Higgs mass. Note that we have chosen the stop masses such that $m_{h_1}=126\gev$ requires a tree-level contribution larger than $M_Z$. In the parameter space where both regions overlap we, thus, simultaneously obtain a solution to the gravitino problem and an enhancement of the tree-level Higgs mass. 

If we increase the value of $\tan\beta$ compared to Fig.~\ref{fig:regions}, the blue and orange regions start moving apart. Consistent with our considerations above, the overlap vanishes for $\tan\beta \gtrsim 1.9$. This observation holds independent of the neutralino mass considered.

\section{Implications for Higgs searches at the LHC}
\label{sec:HiggsLHC}
In this section we shall study the collider implications of the NMSSM with a gravitino LSP. The efficient neutralino annihilation required in this scenario narrows the viable parameter space of the model. It is to clarify, whether any robust predictions arise.

Due to the enhanced tree-level Higgs mass, the NMSSM is consistent with relatively light superpartner spectra. Searches for supersymmetry at LHC-14 therefore look particular promising. We refrain from considering these searches in detail, as they are subject to a strong model-dependence. In particular, they are very sensitive to the masses of the colored sparticles which are otherwise not important for the scenario described here. Direct searches for neutralinos /charginos in the leptonic channels have not yet reached the sensitivity to probe lightest neutralino masses above $100\gev$ unless additional light sleptons appear in the decay chains~\cite{ATLAS:2013rla}. 

On the other hand, clear predictions concerning the Higgs sector of the NMSSM can be made. Therefore, we focus on Higgs searches at the LHC. While the decay rates of the new boson observed at $m_h=126\gev$ are in good agreement with the Standard Model expectations, they still leave considerable room for new physics. We will study, whether the NMSSM with a gravitino LSP predicts any deviations from the branching fractions of the SM Higgs. The LHC has also performed various searches, aiming to probe the existence of an extended Higgs sector. We are going to investigate the existing constraints as well as the the most promising search strategies for the singlet-like and heavy Higgs bosons.

\subsection{Analysis procedure}

Our analysis aims at identifying correlations between the proposed solution to the gravitino problem and Higgs physics at the LHC. In order to search for generic predictions, we have created a benchmark sample of NMSSM points consistent with a gravitino LSP. We restrict our attention to the case where $h_1$ is the SM-like Higgs. The case of a SM-like $h_2$ is tightly constrained by LEP and gives rise to a qualitatively different phenomenology. It is, therefore, left for future work.

For the benchmark sample, the input parameters at the low scale were chosen randomly within the intervals
\begin{align}\label{eq:boundaries}
 \lambda &= 0.5\dots 0.75\;, &\kappa &=-0.5\dots 0.5\;, & \tan\beta &= 1.3\dots 5\;,\nonumber\\
 \mu &= 100\dots 400\gev\;, & m_a &= 200\dots 800\gev\;,& M_1&=250\dots 500\gev\;,\nonumber\\
 A_\kappa &=-200\dots 200\gev\;,
\end{align}
where we traded $A_\lambda$ against $m_a$ as input parameter (cf.~\eqref{eq:pseudomass}). The remaining gaugino masses are determined by gaugino mass unification at the high scale which implies $M_1:M_2:M_3=1:2:6$. The sfermion masses and trilinear couplings were fixed as $m_0=A_0=1.5\tev$. The resulting particle spectrum, Higgs decay rates and the neutralino freeze-out density were determined with NMSSMTools. We have applied the following constraints
\begin{itemize}
 \item consistency with a gravitino LSP, i.e. $\Omega_\chi h^2$ satisfies~\eqref{eq:limitomega},
 \item $\lambda$, $\kappa$ satisfy the perturbativity bound~\eqref{eq:perturbativitybound},
 \item $m_{h_1} = 122-130\gev$,
 \item $m_{a_1} > m_{h_1}/2$.
 \item constraints from flavor/ precision observables.\footnote{Concerning flavor/ precision observables we apply the constraints implemented in NMSSMTools.}
\end{itemize}
A rather loose bound on the Higgs mass was chosen as one can easily change $m_{h_1}$ by a few GeV through varying $A_0$ and $m_0$ within a reasonable range. This would otherwise not affect the scenario. The constraint on $m_{a_1}$ was applied to avoid the decay $h_1\rightarrow a_1 a_1$ which would otherwise upset the Higgs branching ratios.

\subsection{Branching Ratios of the SM-like Higgs}

To study the decay rates of $h_1$, which we assume to be the state observed by ATLAS and CMS, it is instructive to use the basis $(H,h,h_s)$ introduced in section~\ref{sec:notation} (see Appendix~\ref{sec:higgsmatrix} for the mass matrix in this basis). To be consistent with the observed decay pattern of the Higgs, the state $h_1$ should be dominated by $h$ which couples to SM matter exactly as the SM Higgs. Mixing with the singlet and the heavy doublet Higgs tends to suppress the interactions of $h_1$ with $W,Z$ bosons and up-type fermions. Depending on the relative sign of the $h$ and $H$ components in $h_1$, the coupling to down-type fermions can be enhanced or suppressed compared to the SM. Note that this is due to the $\tan\beta$-enhanced coupling of $H$ to down-type fermions. The singlet Higgs $h_s$ couples to charginos with the full strength of $\lambda$ giving rise to a loop-induced coupling to photons. Therefore, the singlet component can increase the partial width of $h_1$ into photon pairs~\cite{SchmidtHoberg:2012yy}.

The signal strength of the Higgs $\mu_{xx}$ obtained by combining the LHC and Tevatron searches in various channels can be found in Tab.~\ref{tab:higgsbranching}. Here we defined
\begin{equation}
 \mu_{xx} ~=~ \frac{\sigma_h\times \text{Br}(h\rightarrow xx)}{\sigma_{h_{\text{SM}}}\times \text{Br}(h_{\text{SM}}\rightarrow xx)}
\end{equation}
as the product of production cross section $\sigma_h$ and branching ratio $\text{Br}(h\rightarrow xx)$ normalized to the corresponding SM quantities. Within experimental errors, the measured signal strengths are in good agreement with the SM predictions.

\begin{table}[t]
\centering
 \begin{tabular}[h]{|c|c|c|c|c|}
\hline
$\mu_{\gamma\gamma}$ & $\mu_{ZZ}$ & $\mu_{WW}$ & $\mu_{bb}$ & $\mu_{\tau\tau}$ \\
\hline
$1.18 \pm 0.19$ & $1.06\pm 0.23$ & $0.79 \pm 0.22$ & $1.01\pm 0.44$ & $ 0.97 \pm 0.31$\\
\hline
\end{tabular}
\caption{\footnotesize{Higgs signal strength and $1\sigma$ statistical error in different channels as observed by ATLAS, CMS, CDF and D0~\cite{Ellis:2013lra}.}}
\label{tab:higgsbranching}
\end{table}

\begin{figure}[t]
\begin{center}
\includegraphics[height=5.5cm]{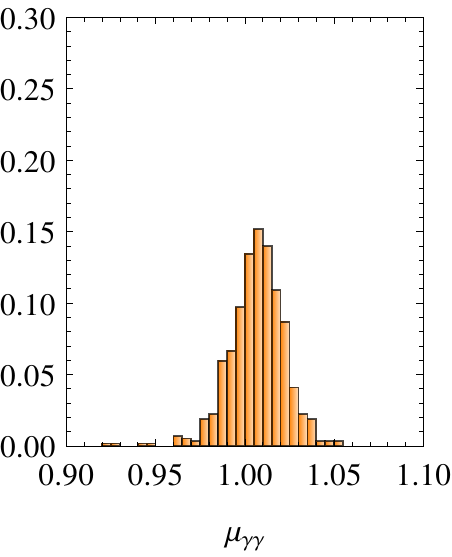}
\hspace{6mm}
\includegraphics[height=5.5cm]{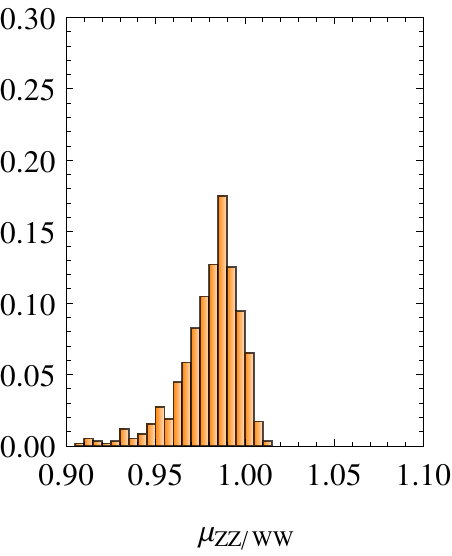}
\hspace{6mm}
\includegraphics[height=5.5cm]{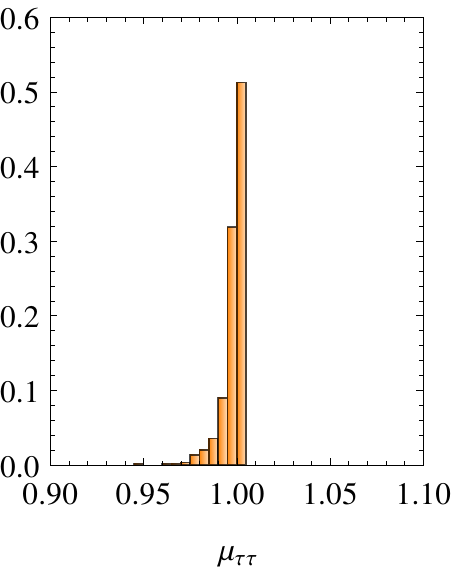}
\end{center}
\caption{\footnotesize{Distribution of $h_1$ signal strengths (normalized to the SM) within the benchmark sample. 
}}
\label{fig:higgsbranching}
\end{figure}

In Fig.~\ref{fig:higgsbranching}, we depict the distribution of Higgs signal strength obtained in our benchmark sample.\footnote{The distribution in the $\bar{b}b$-channel is not shown separately, it resembles the distribution in the $\tau^+\tau^-$-channel.} It can be seen that for all benchmark points $\mu_{xx}$ is very close to the SM prediction which implies that $h_1$ is always strongly dominated by its SM component. Singlet-doublet mixing is suppressed as it would reduce the mass of $h_1$. Further, efficient neutralino annihilation typically requires $m_H\sim m_a > 2\,m_t$ (see section~\ref{sec:channel}) implying the decoupling of the heavier MSSM scalar. As a general trend, the signal strength is very slightly reduced in the $ZZ/WW$ and $\tau^+\tau^-$-channels which results from a smaller production cross section. In the $\gamma\gamma$-channel, there is a competing effect from the enhanced partial decay rate arising from the chargino-loop. The overall signal strength $\mu_{\gamma\gamma}$ can be either reduced or enhanced. 

Even at LHC-14, it will be very challenging to distinguish $h_1$ from the SM Higgs as its signal strength deviates by at most $5\%$ from the SM in all channels. We find that if we drop the requirement of gaugino mass unification, an enhancement of $\mu_{\gamma\gamma}$ by $\sim10\%$ can be achieved for a relatively light wino, similar as in~\cite{SchmidtHoberg:2012ip}. Note also, that if we drop the perturbativity bound~\eqref{eq:perturbativitybound}, larger deviations from the SM branching ratios occur.\footnote{For $\lambda\lesssim 0.8$ Landau poles in the RGE running of $\lambda$ can still be avoided by the inclusion of extra matter in complete SU(5) multiplets~\cite{Masip:1998jc}. We find that for this case deviations of the signal strengths at the level of $\mathcal{O}(10-20\%)$ can occur.} The most promising observable in the scenario discussed here is the ratio $\mu_{ZZ}/\mu_{\gamma\gamma}$ which factors out uncertainties in the Higgs production cross section and is always enhanced compared to the SM.

\subsection{Searches for Higgs bosons}

The possibility to probe the extended Higgs sector of the NMSSM has recently been discussed by various authors~\cite{Kang:2013rj,Christensen:2013dra,Barbieri:2013hxa,Ellwanger:2013ova,Barbieri:2013nka}. In this work, we consider a specific region of the NMSSM parameter space where the neutralino undergoes very efficient pair-annihilations. This allows us to narrow down the promising search channels compared to the general NMSSM. We will first turn to the MSSM-like Higgs bosons and then discuss the singlet-like states.

\begin{figure}[t]
\begin{center}
\includegraphics[height=5.5cm]{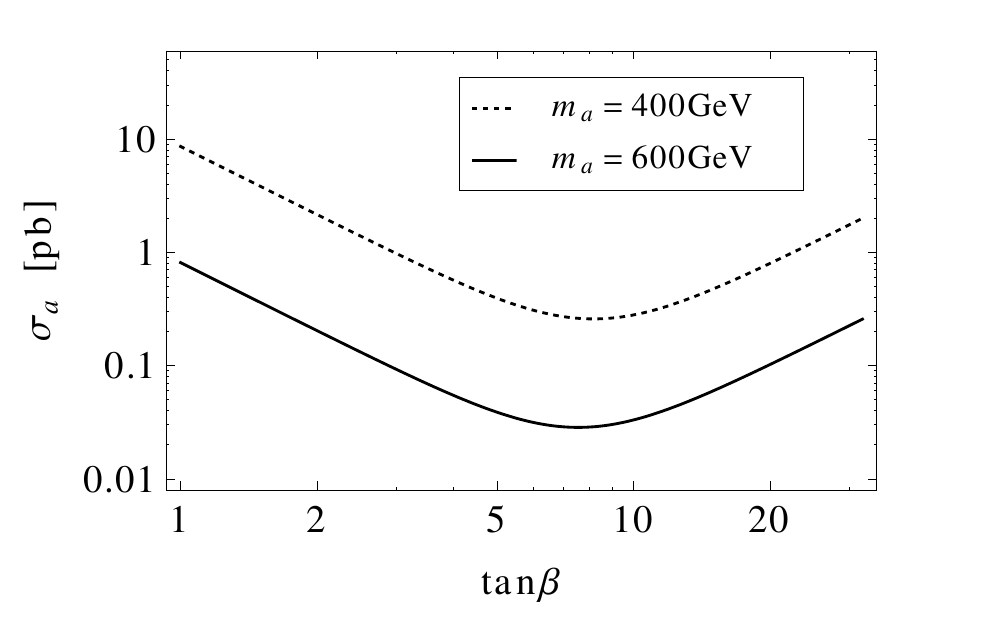}
\end{center}
\caption{\footnotesize{Production cross section of a MSSM-like pseudoscalar Higgs at LHC with $\sqrt{s}=8\tev$ for two different masses as a function of $\tan\beta$.}}
\label{fig:prodcross}
\end{figure}

In the MSSM, the production of the heavy Higgs bosons is typically suppressed. Their coupling to top quarks is reduced by a factor of $\tan\beta$, rendering the gluon fusion of $H$ and $a$ less efficient. Sizeable cross sections can only be obtained at large $\tan\beta$ where Higgs production in bottom quark fusion and gluon fusion via a bottom-loop becomes efficient. The production cross section $\sigma_a$ of a MSSM-like pseudoscalar as a function of $\tan\beta$ is depicted in Fig.~\ref{fig:prodcross}. Here we have used the tools HIGLU~\cite{Spira:1995mt} and bbh@nnlo~\cite{Harlander:2003ai} to determine the dominant contributions to $\sigma_a$ at next-to-next-to-leading order.\footnote{Here we neglected possible contributions to the cross sections induced by the superpartners.} The production cross section $\sigma_H$ for the CP even Higgs is slightly smaller than $\sigma_a$, but has a similar scaling with respect to $\tan\beta$.

In the regime of large $\tan\beta$, strong lower bounds on the mass of the MSSM pseudoscalar arise from searches for $\tau^+\tau^-$ final states performed by ATLAS and CMS~\cite{Chatrchyan:2012vp,Aad:2012cfr}. The constraints become considerably weaker at intermediate values of $\tan\beta\sim 10$ where the cross section $\sigma_a$ has a minimum. Small values of $\tan\beta$ can hardly be accessed in the MSSM due to the constraints on the light Higgs mass.

Turning to the NMSSM, the window at low $\tan\beta$ is reopened due to the new tree-level contributions to $m_{h_1}$. If we require the NMSSM to be consistent with a gravitino LSP we even found an upper limit $\tan\beta < 1.9$ (see section~\ref{sec:limittan}). This implies that the production cross sections $\sigma_{a,H}$ are expected to be quite sizeable. At low $\tan\beta$, the heavy Higgs bosons would -- if kinematically allowed -- dominantly decay into top-pairs rather than down-type fermions. The Higgs groups of ATLAS and CMS have not yet performed a dedicated search for Higgs bosons with subsequent decay to top quarks. However, the Exotics (ATLAS) and B2G (CMS) groups have recently provided constraints on the production cross section of general top-top resonances~\cite{ATLAS:2013-052,CMS:12-006} which are also applicable to the case of a Higgs boson (see also~\cite{Arbey:2013jla,Djouadi:2013vqa}).

\begin{figure}[t]
\begin{center}
\includegraphics[height=5.0cm]{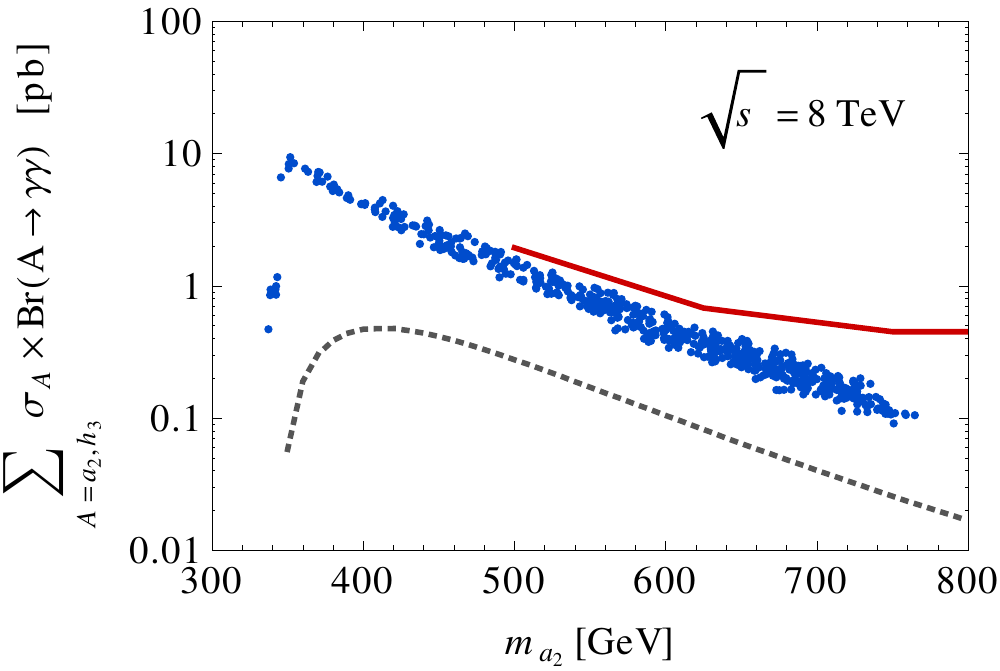}
\hspace{5mm}
\includegraphics[height=5.0cm]{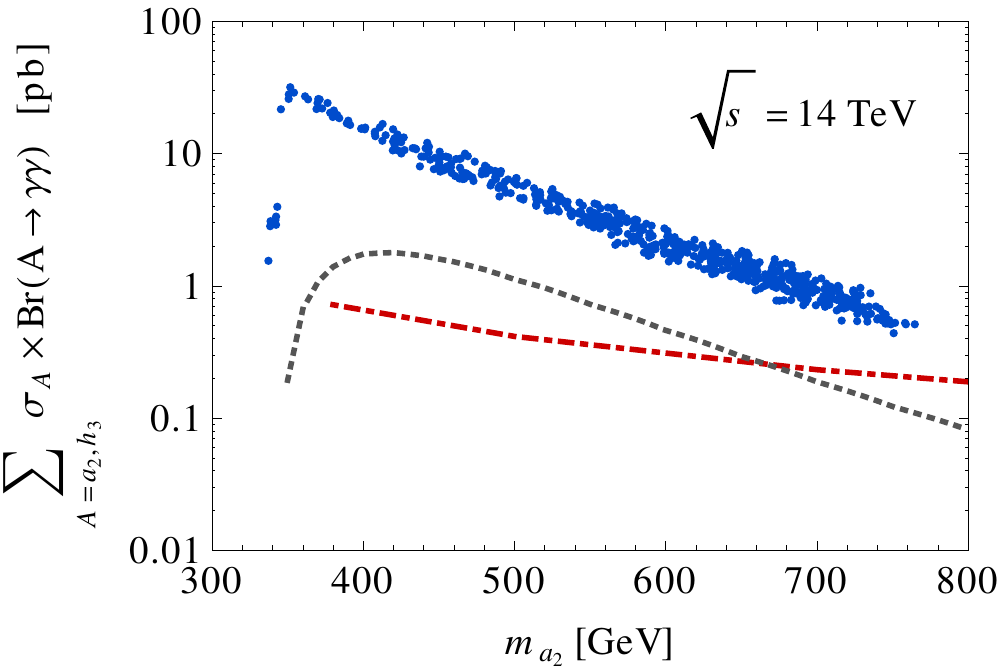}
\end{center}
\caption{\footnotesize{Production cross section of the heavy MSSM-like Higgs bosons weighted by their branching ratio into top-pairs in our benchmark sample (blue points) at LHC-8 (left panel) and LHC-14 (right panel). For comparison the same quantity is shown for a SM Higgs with mass $m_{a_2}$ (gray dashed contour). The current limit from the CMS search for top-top resonances at LHC-8 is shown as the red solid line (left panel), the expected sensitivity of LHC-14 at $200\:\text{fb}^{-1}$ as the red dash-dotted line (right panel).}}
\label{fig:topsearch}
\end{figure}

In our benchmark sample, $a_2$ and $h_3$ are the heavy MSSM-like Higgs fields. Mixing effects with the singlet states are typically suppressed. As $a_2$ and $h_3$ are usually very close in mass, they cannot be resolved as single particles in a low resolution channel like $\bar{t}t$ and contribute to the same resonance. We have used HIGLU and bbh@nnlo to calculate the production cross sections $\sigma_{a_2}$ and $\sigma_{h_3}$ with the effective couplings taken from NMSSMTools. Decay rates were also determined with NMSSMTools. In Fig.~\ref{fig:topsearch}, we depict the heavy Higgs production cross section times branching ratio into top quarks at $\sqrt{s}=8\tev$ and $\sqrt{s}=14\tev$, where we performed the sum over the contributions from $a_2$ and $h_3$. In the left panel, we have also depicted the best current limit from CMS~\cite{CMS:12-006}. In the right panel we provide the estimated sensitivity of LHC-14 at $200\:\text{fb}^{-1}$. This estimate was obtained from the ATLAS projected sensitivity to top-top resonances at $\sqrt{s}=10\tev$~\cite{ATLAS:2010-008} by a simple rescaling.\footnote{In~\cite{ATLAS:2010-008} the ATLAS sensitivity to top-top resonances was estimated for a center-of-mass energy of $10\tev$ and a luminosity of $200\:\text{pb}^{-1}$. To correct for the change in center of mass energy and luminosity we have multiplied the estimated sensitivity by $\sqrt{2/1000}$. The factor $2$ accounts for the cross section of the dominant background processes being enhanced by a factor $\sim 2$ if one raises the center-of-mass energy from $10$ to $14\tev$. The factor $1000$ arises as we assumed a luminosity of $200\:\text{fb}^{-1}$ compared to $200\:\text{pb}^{-1}$ in~\cite{ATLAS:2010-008}.} Invariant masses $m_{tt}<500$ of the top-top system are not covered in~\cite{ATLAS:2010-008} and we assumed a scaling of the sensitivity with $1/m_{tt}^2$ as suggested in~\cite{Djouadi:2013vqa}.
It can be seen that all benchmark points are very close to the current limit on top-top resonances. Some of the parameter points at $m_{a_2}\sim 400\gev$ could already be tested if CMS would extend its analysis range to lower masses. The prospects for the detection of the heavy Higgs bosons at LHC-14 are very promising: the entire parameter space spanned by the benchmark sample will be probed. This prediction is a direct consequence of requiring the NMSSM to be consistent with gravitino dark matter. It follows from the low $\tan\beta$ required in this scenario (see section~\ref{sec:limittan}) and holds unless for very large masses $m_{a_2}\gg 1\tev$. We excluded such large $m_{a_2}$ in our benchmark sample as they would correspond to $\mu\gg 500\gev$ which is disfavored by electroweak naturalness. A few benchmark points around or below the top threshold cannot be tested in the $\bar{t}t$-channel. These parameter points do not escape detection, they will be probed in the $\tau^+\tau^-$ or -- as we shall see -- in the $\gamma\gamma$-channel.\footnote{The mass region below the top threshold is also covered by searches for $h_3$ in the $ZZ$-channel.} 

Turning to the singlet-like states $h_2$ and $a_1$, their production cross section is generically suppressed. At tree-level, they only couple to SM matter via their subleading doublet components. The $\gamma\gamma$-channel is most promising in the search for the singlet-like states. In particular, the pseudoscalar $a_1$ typically has a large branching ratio into photons. This decay mode is mediated by higgsinos in the loop which couple to $a_1$ with the full strength of $\lambda$. Indeed, it can be of similar magnitude as the competing decay mode into $\bar{b}b$-pairs. The latter is a tree-level process, but it is suppressed by the small mixing angle between singlet and doublet pseudoscalar. The heavy MSSM-like Higgs fields can also be searched for in the $\gamma\gamma$-channel. However, only if $\bar{t}t$ final states are kinematically inaccessible, the branching fraction into photon pairs is sufficiently large to be detectable.

\begin{figure}[t]
\begin{center}
\includegraphics[height=5.0cm]{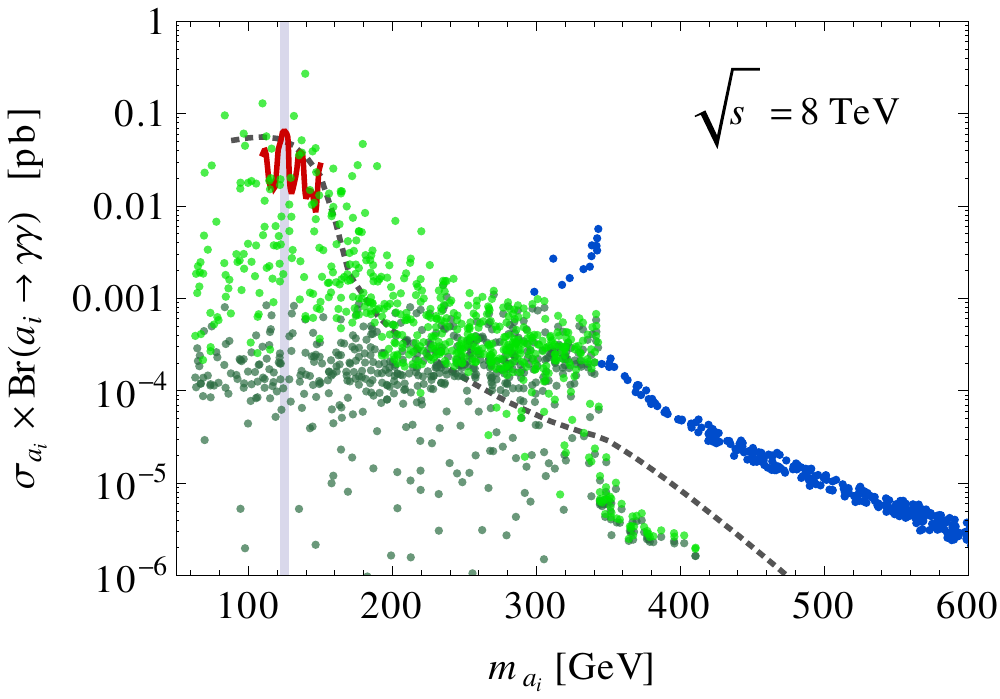}
\hspace{5mm}
\includegraphics[height=5.0cm]{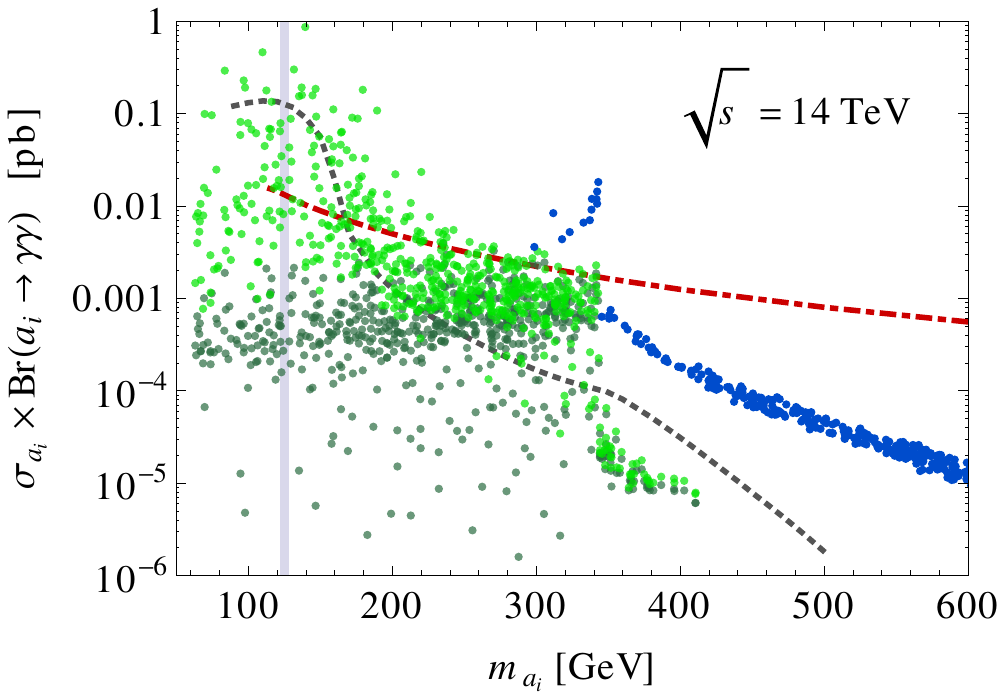}
\end{center}
\caption{\footnotesize{Production cross section times diphoton decay rate of the MSSM-like pseudoscalar $a_2$ (blue points) and the singlet-like pseudoscalar $a_1$ (green points) in our benchmark sample at LHC-8 (left panel) and LHC-14 (right panel). The dark green points refer to the case, where only the direct production of $a_1$ is considered, while the light green points refer to the case where also the indirect production of $a_1$ via decay of the heavier Higgs bosons is taken into account. For comparison, the cross section times diphoton decay rate of a Standard Model Higgs is also shown (gray dashed contour). The current limit in the $\gamma\gamma$-channel from the CMS MVA analysis is indicated by the red solid line (left panel), the expected sensitivity of LHC-14 at $200\:\text{fb}^{-1}$ by the red dash-dotted line (right panel). The shaded band shows the mass range, in which the diphoton decay of $a_1$ would be misidentified as part of the SM Higgs signal.}}
\label{fig:gammasearch}
\end{figure}

To summarize, the singlet-like pseudoscalar $a_1$ has a large branching fraction into photons but its production cross section is suppressed. For the heavy MSSM-like states, the situation is exactly opposite. This suggests to look for the indirect production of $a_1$ via decay of the heavier Higgs fields with the subsequent decay of $a_1$ into photons. Possible decay sequences driven by the couplings $\lambda$ and $\kappa$ include
\begin{equation}\label{eq:chaindecay}
 a_2 \rightarrow h_2 Z \rightarrow a_1 a_1 Z\;, \quad  h_3 \rightarrow h_1 h_2 \rightarrow h_1 a_1 a_1\;.
\end{equation}
If these processes are kinematically allowed the branching ratios $\text{Br}(h_3,a_2\rightarrow a_1 + X)$ with arbitrary $X$ typically reach 1-10\%. Therefore, the indirect production of $a_1$ can be much more efficient than the direct production and could give rise to a feature in the diphoton invariant mass distribution. In Fig.~\ref{fig:gammasearch} we provide the cross section times diphoton branching fraction for both pseudoscalars in our benchmark sample (at $\sqrt{s}=8\tev$ and $\sqrt{s}=14\tev$). For the singlet-like state $a_1$ we separately depict the cases with and without taking into account the indirect production via heavy Higgs decays.
Also shown are the current constraints in the $\gamma\gamma$-channel from the CMS MVA analysis~\cite{CMS:ril} and the LHC-14 projected sensitivity~\cite{Pieri:2006bm}. In the regime of invariant masses $m_{\gamma\gamma} > 150\gev$ not covered in~\cite{Pieri:2006bm}, we assumed a scaling of the sensitivity with $1/m_{\gamma\gamma}^2$.\footnote{This type of scaling is suggested if one compares the sensitivity of the LHC experiments in the range of invariant masses $m_{\gamma\gamma}=100-150\gev$ covered by the Higgs searches and the range $m_{\gamma\gamma}> 500\gev$ covered by the searches for extra dimensions~\cite{Chatrchyan:2011fq,Aad:2012cy}.}
Considering only the direct production of $a_1$ via gluon fusion, it would be difficult to find a signal for the singlet-like pseudoscalar -- even at LHC-14. If we include its indirect production by heavy Higgs decay, the chances to observe an excess in the $\gamma\gamma$-channel are considerably larger, especially if $m_{a_1}\leq 200\gev$. We should note that the LHC-8 limit / LHC-14 projected sensitivity is not strictly applicable if $a_1$ is produced by decays as this would affect the  kinematics and event characteristics. Indeed, one might even be able to increase the sensitivity to such events by triggering on additional decay products obtained in the decays~\eqref{eq:chaindecay} like additional $\bar{b}b$ or lepton pairs.

Let us mention that if, by chance, $m_{a_1}=122 -128\gev$, the diphoton decay of $a_1$ could be misidentified as part of the SM Higgs signal. This could fake an enhancement of the branching ratio $\mu_{\gamma\gamma}$~\cite{Munir:2013wka}. 

Turning to the MSSM-like pseudoscalar, we find that the benchmark points with $m_{a_2} < 2\,m_t$ can be tested in the diphoton channel at LHC-14. This is very appealing as it implies that the whole range of $m_{a_2}$ is covered either in the $\gamma\gamma$- or the $\bar{t}t$-channel. Our scenario is thus completely testable at LHC-14.

\section{Conclusion}

We have presented a scenario with weak-scale gravitino dark matter in the simplest NMSSM model with soft parameters arising from gravity-mediated supersymmetry breaking, including gaugino mass unification. We exploited the coupling $\lambda$ to enhance the tree-level Higgs mass such that the required loop corrections are achieved with superpartner masses that remain in the TeV regime.
Simultaneously, the same coupling gives rise to very efficient annihilation of a mixed higgsino/singlino NLSP
via the $\lambda S H_u H_d$ term in the superpotential.
The dominant process is s-channel pair-annihilation via a pseudoscalar into a pair of top quarks as shown in Fig.~\ref{fig:feynman}.
Analytic approximations for the corresponding annihilation cross section~\eqref{eq:approximation} are provided. 
If the pseudoscalar mass falls in the vicinity of the resonance --requiring a relatively mild tuning at the level of $10\%$-- the neutralino density at freeze-out is sufficiently suppressed to satisfy the severe cosmological bounds even for lifetimes as large as $\tau\sim 10^{10}\s$. 
For the case under study we could cast these bounds in a surprisingly simple form~\eqref{eq:limitomega}.
Thermally produced gravitinos account for the observed dark matter in the universe whereby in our scenario the
required  reheating temperature is sufficiently high to produce the visible matter in the universe via standard thermal leptogenesis.

Our scenario constraints the NMSSM parameters, cf.~Fig.~\ref{fig:regions}, and $1.3 < \tan{\beta} \lesssim 1.9$ which allowed us to identify the decisive search channels at the LHC.
In parts of the parameter space a singlet-like pseudoscalar is within reach in the diphoton channel at LHC-14.
Most importantly, our scenario predicts the existence of a pseudoscalar Higgs with dominant decay into top quarks. This state virtually always has a production cross section sufficiently large to be detected in the upcoming searches for top-top resonances at LHC-14.
If its mass is below the top threshold, its existence will be tested in the diphoton channel. 
Taken together, we obtain our key finding: the NMSSM with gravitino dark matter and sparticle masses from ordinary gravity mediation can be tested at LHC-14.

Altogether, we have found a fascinating example for the interplay between cosmology and collider physics. 
In a theory motivated by simplicity and the recent discovery of the Higgs boson, the gravitino forms the observed dark matter in the universe and a consistent cosmology becomes possible. 
We find it remarkable that 
even though the gravitino evades any direct detection, our scenario is completely testable at the LHC via its implications on the Higgs sector.
By taking the cosmology of our universe into account, falsifiable predictions for the LHC became feasible.

\section*{Acknowledgments}
We would like to thank Michael Ratz for useful discussions and Jim Cline for correspondence (see main text). 
JH was supported by the German Research Foundation (DFG) via the
Junior Research Group ``SUSY Phenomenology'' within the Collaborative
Research Centre 676 ``Particles, Strings and the Early Universe''.
JH would like to thank the German Academy of Science for support through the Leopoldina Fellowship Programme 
grant LPDS 2012-14.

\appendix

\section{Mass matrices}

In the following, we provide the neutralino and Higgs mass matrices at tree-level. We mainly follow the conventions of~\cite{Ellwanger:2009dp}.

\subsection{Neutralino mass matrix}\label{sec:neutralinomatrix}
The neutralino mass matrix $\mathcal{M}_\chi$ in the basis $(-\I\, \widetilde{B}, -\I\, \widetilde{W} , \widetilde{h_d}, \widetilde{h_u},\widetilde{s})$ reads
\begin{equation}
 \mathcal{M}_\chi ~=~ \begin{pmatrix}
                   M_1 & 0 & -\frac{g_1 v_d }{\sqrt{2}} & \frac{g_1 v_u }{\sqrt{2}} & 0 \\
                   0  & M_2 & \frac{g_2 v_d }{\sqrt{2}}  & -\frac{g_2 v_u }{\sqrt{2}} & 0 \\
                  -\frac{g_1 v_d }{\sqrt{2}} & \frac{g_2 v_d }{\sqrt{2}} & 0 & -\mu & \lambda v_u\\
                   \frac{g_1 v_u }{\sqrt{2}} & -\frac{g_2 v_u }{\sqrt{2}} &-\mu & 0& \lambda v_d\\
                   0 & 0 & \lambda v_u & \lambda v_d & \frac{2\,\kappa}{\lambda}\mu
                    \end{pmatrix}\;,
\end{equation}
where we have used
\begin{equation}
 \mu ~=~ \lambda\, v_s\;.
\end{equation}

\subsection{Higgs mass matrices}\label{sec:higgsmatrix}

The tree-level mass matrix $\mathcal{M}_{A}^2$ of the CP odd Higgs fields in the basis $(a,a_s)$ reads
\begin{equation}
 \mathcal{M}_{A}^2 ~=~ \begin{pmatrix}
                     m_a^2 & m_{aa_s}^2\\
                     m_{aa_s}^2 & m_{a_s}^2
                     \end{pmatrix}
\end{equation}
with
\begin{equation}\label{eq:pseudomass}
m_a^2 ~=~ \frac{2\,\mu\,(\lambda A_\lambda + \kappa \mu)}{\lambda \sin{2\beta}}\;.
\end{equation}
The remaining entries expressed in terms of $m_a$ are given as
\begin{subequations}
\begin{eqnarray}
m_{a_s}^2&=& \lambda^2 v^2 \,\frac{m_a^2 \sin^2{2\beta}}{4\mu^2} -3\,\kappa\mu\,\frac{A_\kappa}{\lambda}+ \frac{3}{2}\lambda \kappa\, v^2 \,\sin{2\beta}\;,\\
m_{aa_s}^2&=& \lambda v\,\frac{m_a^2 \sin{2\beta}}{2\mu}- 3\kappa \, v\mu\;.
\end{eqnarray} 
\end{subequations}
The mass matrix $\mathcal{M}_{H}^2$ of the CP even Higgs fields in the basis $(H,h,h_s)$ reads
\begin{equation}
 \mathcal{M}_{H}^2 ~=~ \begin{pmatrix}
                      m_H^2 & m_{Hh}^2 & m_{Hh_s}^2\\
                      m_{Hh}^2 & m_h^2 & m_{hh_s}^2\\
                      m_{Hh_s}^2 & m_{hh_s}^2 & m_{h_s}^2
                     \end{pmatrix}
\end{equation}
with
\begin{subequations}
\begin{eqnarray}
m_H^2&=& m_a^2 + (M_Z^2-\lambda^2 v^2) \sin^2{2\beta}\;, \\
m_h^2&=& M_Z^2 \cos^2{2\beta} + \lambda^2 v^2 \sin^2{2\beta}\;, \\
m_{h_s}^2&=& \lambda^2 v^2 \,\frac{m_a^2 \sin^2{2\beta}}{4\mu^2} + \kappa\mu \,\frac{4\kappa\mu+\lambda A_\kappa}{\lambda^2}- \frac{1}{2}\lambda \kappa\, v^2 \,\sin{2\beta}\;, \\
m_{Hh}^2&=& \frac{1}{2}(M_Z^2-\lambda^2 v^2) \sin{4\beta} \;, \\
m_{Hh_s}^2&=& \kappa v\,\mu \cos{2\beta} + \frac{\lambda v\, m_a^2}{4\,\mu^2}\,\sin{4\beta}\;,\\
m_{hh_s}^2&=& (2\lambda - \kappa \sin{2\beta})\,v\mu - \lambda v\,\frac{m_a^2 \sin^2{2\beta}}{2\mu}\;.
\end{eqnarray} 
\end{subequations}

\bibliography{gravitino}
\bibliographystyle{utphys2}

\end{document}